\documentclass{jpconf}
\usepackage{graphicx}
\usepackage{amsmath}
\usepackage{amssymb}
\usepackage{lmodern}

\begin{document}

\title{Influence of the random walk  finite step on the first-passage probability}
\author{Olga Klimenkova$^{1, 2}$, Anton Menshutin$^{2}$, Lev Shchur$^{1, 2, 3}$}
\address{$^1$ National Research University Higher School of Economics, 101000 Moscow, Russia}
\address{$^2$ Science Center in Chernogolovka, 142432 Chernogolovka, Russia} 
\address{$^3$ Landau Institute for Theoretical Physics, 142432 Chernogolovka, Russia}

\begin{abstract}
A well known connection between first-passage probability of random walk and distribution of electrical
potential described by Laplace equation is studied. We simulate random walk in the plane numerically as a discrete time process
with fixed step length. We measure first-passage probability to touch the absorbing sphere of radius $R$ in 2D.
 We found a regular deviation of the first-passage probability  from the exact function, which we attribute to the finiteness of the random walk step.
\end{abstract}

\section{Introduction}
Connection between stochastic processes and boundary value problems is well known. A Brownian motion is a classical
example of continuous time stochastic process and it was shown by Einstein~\cite{Einstein} that a collective motion of
many Brownian particles is governed by diffusion equation. Let $B$ be a domain in the plane with boundary $\partial B$. Assume 
we have a Brownian particle that starts walking from some position $\boldsymbol{r}_0$, and the boundary $\partial B$ is absorbing which
means that once a walker hits the boundary the walk is terminated. The time-dependent behavior of Brownian particle is described
by diffusion equation:
\begin{equation}
\frac{\partial\rho(\boldsymbol{r},t)}{\partial t} = D\nabla^2\rho (\boldsymbol{r},t)
\label{eq:diffusion}
\end{equation}
with the boundary condition $\rho(\boldsymbol{r},t)=0$, $r \in \partial B$ and initial condition 
$\rho(\boldsymbol{r},t=0)=\delta(\boldsymbol{r}-\boldsymbol{r}_0)$.

If we are interested only in time-independent properties, e.g. first-passage probability, integration over time gives Laplace equation~\cite{Redner}
\begin{equation}
D\nabla^2\varphi(\boldsymbol{r})=-\delta(\boldsymbol{r}-\boldsymbol{r}_0).
\label{eq:Laplace}
\end{equation}
Equation~(\ref{eq:Laplace}) describes distribution of electrical potential $\varphi$ created by a point charge located at 
$\boldsymbol{r}_0$. The absorbing boundary condition in diffusion equation translates to the condition
that a boundary $\partial B$ is grounded, $\varphi(\boldsymbol{r})=0$, $r \in \partial B$.

These connection allows one to analytically investigate Brownian motion by solving either diffusion equation or Laplace equation.

Vice-Versa, these connection could be used to numerically solve some boundary value problems. 
It was shown by Kakutani~\cite{Kakutani} that solution of the equation
\begin{equation}
\nabla ^2 u(x)=0, \; x \in B
\label{eq:nabla}
\end{equation}
with the boundary condition 
\begin{equation}
\lim_{y\rightarrow x} u(y)=g(x),\;  x \in \partial B
\label{eq:cbc-nabla}
\end{equation}
at some point $x$ could be found as an expected value of $g(t)$, $t \in \partial B$ for a random walk starting at $x$, 
and $t$ being its first exit point.

Monte Carlo methods for boundary-value problems have several advantages. They are efficient for estimation of the function 
$u(x)$ at some point $x$.  Monte Carlo simulations also have good performance for complex boundaries like fractals, 
and especially effective in  high dimensional space. In addition, the parallel simulation of multiple random walks is straightforward and easy in implementation.

Special care should be taken while simulating random walks. In the study of the fractals formation by successive 
aggregation of random walks, e.g. diffusion limited aggregation (DLA) model~\cite{Witten-Sanders}, the domain $B$ is not bounded and
diffusing particle is allowed to go infinitely far away. It is known that in 2D an escape probability is zero. In other words, 
all diffusing particles will 
be finally attached to DLA cluster after some time. This time can be infinitely large, and in practice one have to 
halt simulation if random walk goes at some large distance. 
For this purpose killing boundary of some big radius around the DLA cluster is used, 
and these may lead to the explosive growth of cluster in one direction. 
To avoid these unnecessary effect, the killing-free algorithm was proposed~\cite{Killing-free,Free-SSZ}, 
   which allows to take into account the infinite boundary conditions exactly.

Discrete time random walk is easy to simulate but finite step of the random walk produces some bias. 
It was shown in~\cite{ZMC}  that particles undergoing discrete-time steps in three dimensions are 
captured with probability which is different from the probability generated with the infinitesimally short steps, 
and difference does depend on the root-mean square distribution of step length.

In the paper we present results of the simulations of the random-walk in the plane and propose regular form of the first correction to the first-passage probability which is due to the finite step of the random-walk.

\section{Simulation algorithm}

\begin{figure}[ht]
\begin{center}
\includegraphics[scale=0.4]{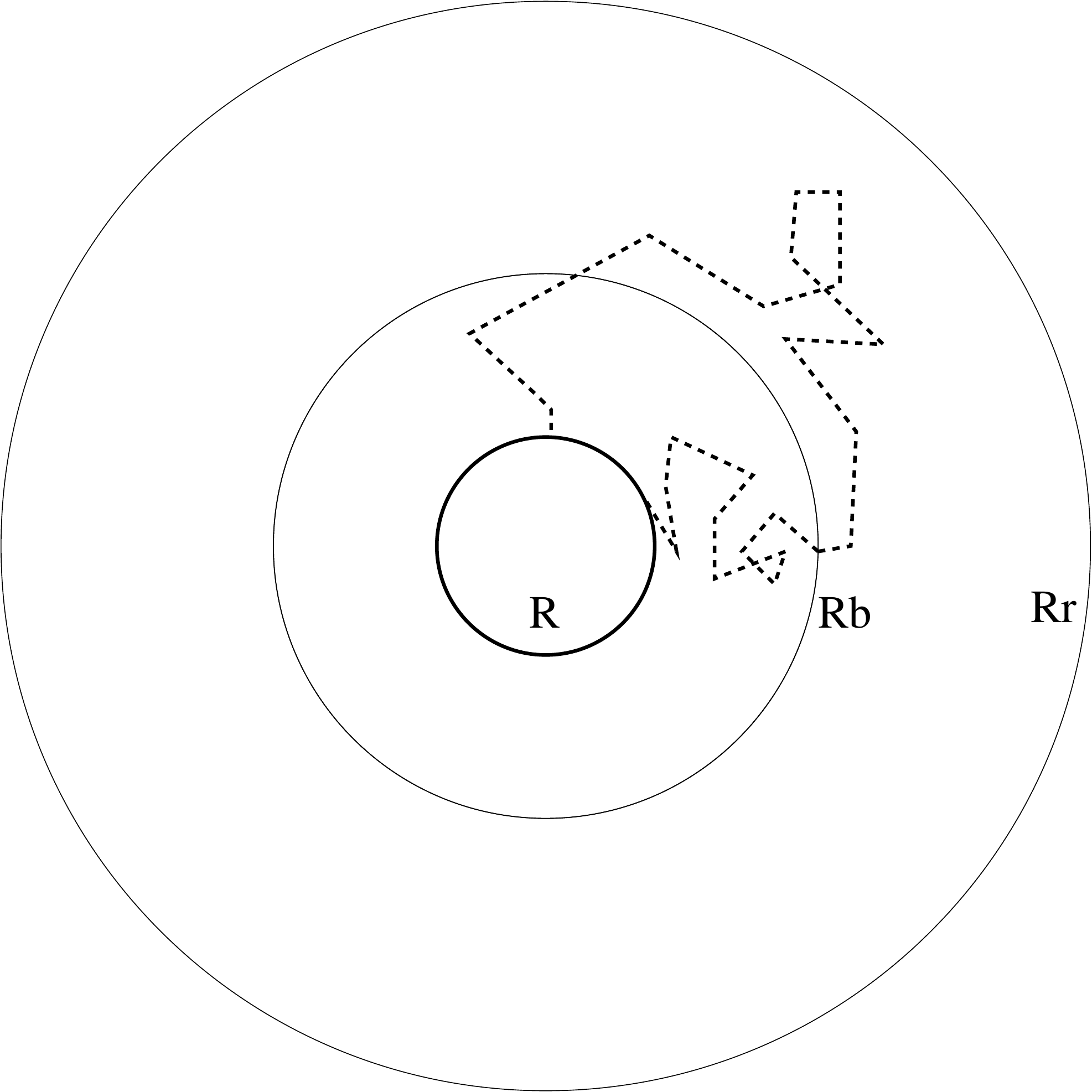}
\caption{Illustration of the random walk in the plane. Random walk of the particle starts at the birth circle of radius $R_b$. Particle terminates while touching the absorbing circle of radius $R$. If particle cross returning circle $R_r$ it is returning back to the birth-circle at the corresponding angle (see Expr.~\ref{eq-return}) and discussion in the text.}
\label{fig:walk-scheme}
\end{center}
\end{figure}

Let us consider particle at the  point $(R_b,0)$ in the plane at the distance $R_b$ from the center of absorbing circle of 
radius $R$ as illustrated in the Figure~\ref{fig:walk-scheme}.

Particle at each time step jumps on the length $\delta$ in random direction. The random walk is symmetric
thus directions of jumps are distributed uniformly. Particle position is continuous variable thereby such a process is the discrete time random walk in continuous space. 
If after making a move particle is inside the circle $R$ at position $(r \cos \phi, r \sin \phi)$, $r<R$
then it is assumed absorbed at position given by angle $\phi$. For simplicity we assume that $(R-r)\sim \delta$ and $\delta \ll 1$ 
so the difference in real absorption position and position after the last jump is negligible.

Probability $P_{exp}(\phi)$ for a particle to be absorbed at the angle $\phi$ is measured numerically. We compare it with 
analytic solution found by solving Laplace equation:
\begin{equation}
P(\phi) = \frac{1}{2\pi} \frac{x^2-1}{x^2-2x\cos \phi +1}
\label{eq-return}
\end{equation}
where $x={R_b}/{R}>1$ (see for details Reference~\cite{Killing-free}).

Expression~(\ref{eq-return}) can be interpreted as a first passage probability for a particle starting at the distance $R_b$ from the center of absorbing circle $R$.
At the same time, it could be used to speed up simulation to deal with fly-away particles problem. The most trivial way to deal with
particles that have gone far from the absorbing circle $R$ is to kill them. But this introduces an error in simulation and should be avoided.
The correct way to solve problem, is to use  expression~(\ref{eq-return}) which allows us to return went away particles back to the simulation region.
In our simulation if particle goes out of $R_r$, $R_r \gg R_b$, it is returned back to $R_b$ with probability~(\ref{eq-return})  and
angle is counted relative to the line connecting the particle position and the center of absorbing circle. 
In order to generate random variable $\phi$ with distribution (\ref{eq-return}) one can use the following mapping
\begin{equation}
\phi = f(u) = 2\arctan \left(\frac{x-1}{x+1}\tan u\frac{\pi }{2}\right),
\label{eq:rng}
\end{equation}
where $u$ is a uniform random number, $u \in [-1, 1]$, and $x={R_b}/{R}>1$. See~\cite{MS-diag} for details.
The killing-free algorithm effectively makes simulation region infinite which results in the correct account of the boundary condition.

Since we study the influence of finite jump length and we measure first-passage probability numerically and compare it with analytical
solution given by Eq.~(\ref{eq-return}), application of the same formula during the simulation should be done with care.
To minimize this influence we must ensure that $(R_b-R) \gg \delta$ and $(R_r-R_b) \gg \delta$.

In the next section, we simulate random walk for different sets of parameters,
 estimate probability distribution $P_{exp}(\phi)$, and study its deviation from the exact result~$P(\phi)$.

\section{Simulation results}

For calculation $P_{exp}(\phi)$, we divide interval  of the possible values of $\phi$ $[-\pi:\pi]$ into 180 bins and count number of hits 
for each bin.
Normalizing results over the total number of random walkers $N$ and over the bin size gives estimation of the hitting probability $P_{exp}(\phi)$.

\begin{figure}[!ht]
\begin{center}
\includegraphics[scale=0.5]{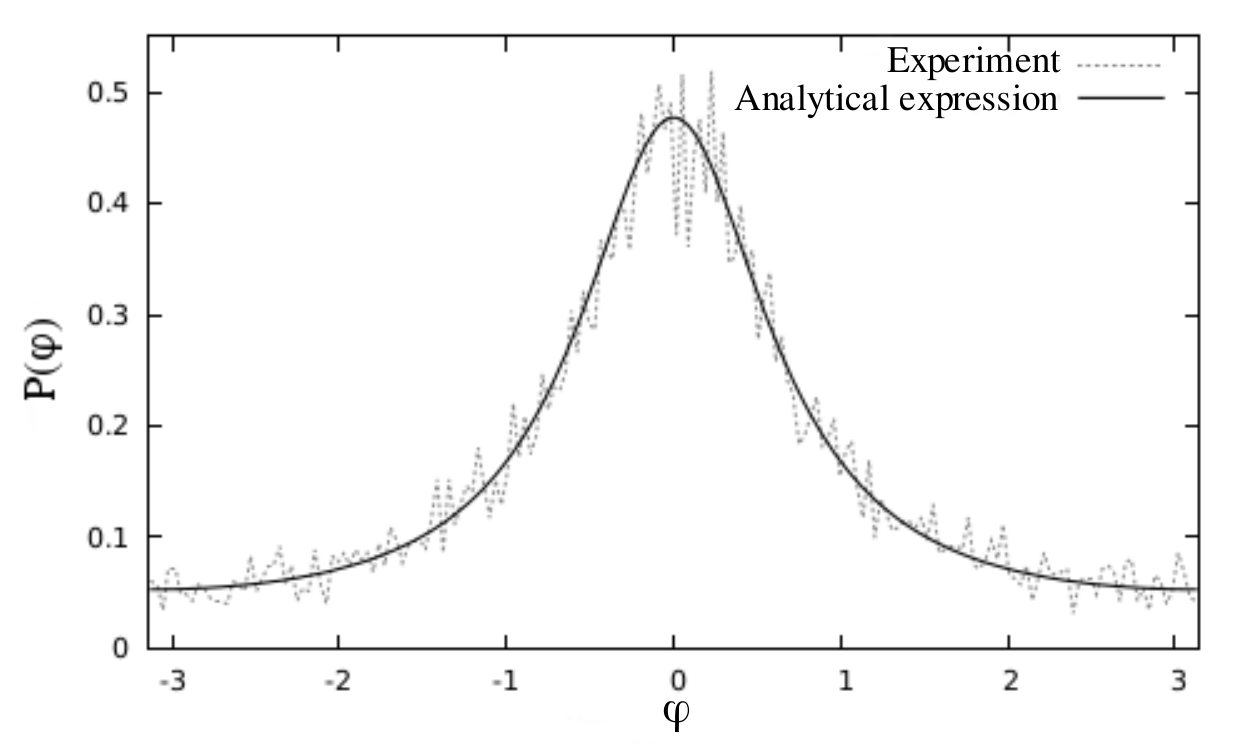}~\includegraphics[scale=0.5]{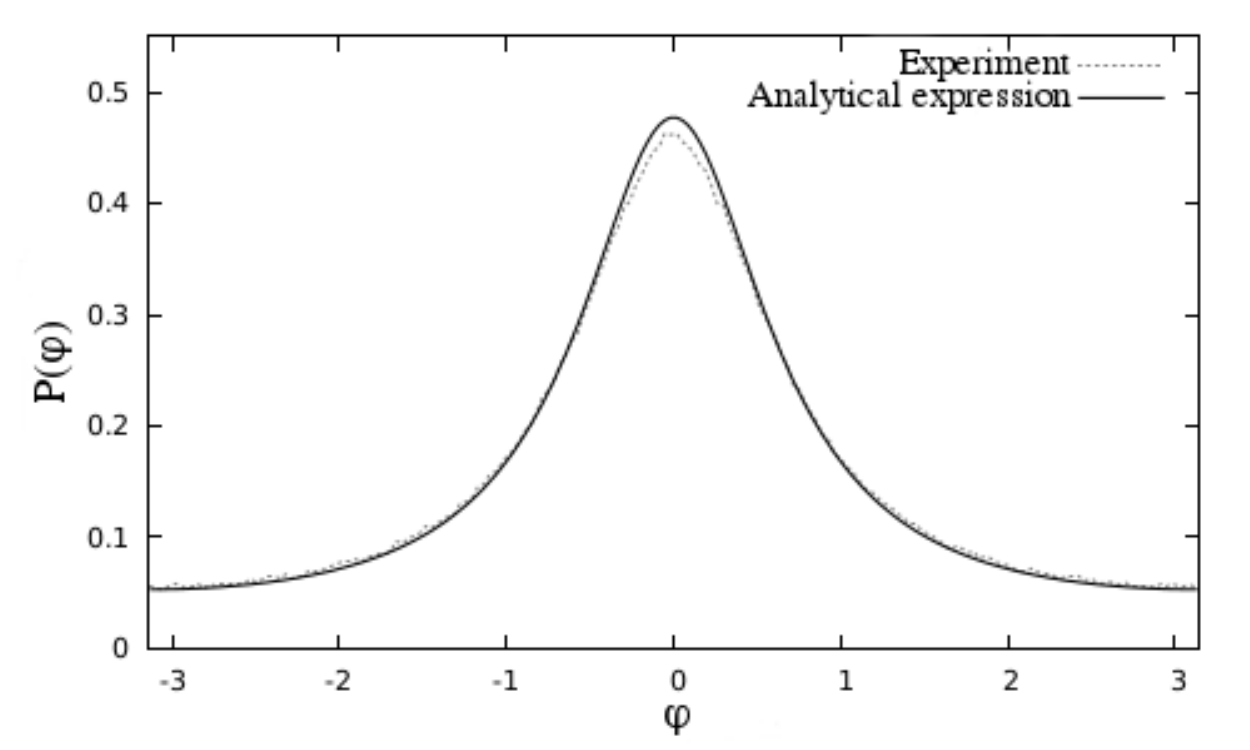}
\caption{Comparison of the estimated probability $P_{exp}$ with the exact probability $P(\phi)$ in Expr.~(\ref{eq-return}). 
    Left: Parameters are $N=10^4$, $R=10$, $R_b=20$, $R_d=200$, $\delta=1$. Right:Parameters are $N=10^6$, $R=10$, $R_b=20$, $R_d=200$, $\delta=1$. }
\label{fig:ex1}
\end{center}
\end{figure}

Comparison of experimental result with analytical is presented in Figure~\ref{fig:ex1}. Data in the left panel is for the relatively small number of runs $N=10^4$ while data in the right panel is for the larger number of runs $10^6$. Fluctuations become not visible on the scale of the figures for the large enough number of runs demonstrating quality of the data. At the same time, it comes clear, that there are some deviations of the measured in experiment probability function $P_{exp}(\phi)$ from the exact one. 

In order to make deviations more visibly pronounced we compute the relative deviation of the estimated probability $P_{exp}$ from the exact probability $P(\phi)$
\begin{equation}
f(\phi)=\frac{P_{exp}(\phi)-P_{exact}(\phi)}{P(\phi)}.
\label{eq:deviation}
\end{equation}

We estimate $f_i(\phi)$ from $N=10^6$  runs, repeat this simulation $M$ times, and calculate an average
\begin{equation}
\left<f(\phi)\right>=\frac{\sum_{i=1}^M f_i(\phi)}{M}.
\label{eq:f-average}
\end{equation}
Standard error of  $f_i(\phi)$ is calculated accordingly 
\begin{equation}
Ef(\phi) = \frac{\sqrt{\sum_{i=1}^M (f_i(\phi) - \left<{f(\phi )}\right>)^2}}{M}.
\label{eq:f-error}
\end{equation}

\begin{figure}[!ht]
\begin{minipage}[h]{0.47\linewidth}
\center{\includegraphics[width=1\linewidth]{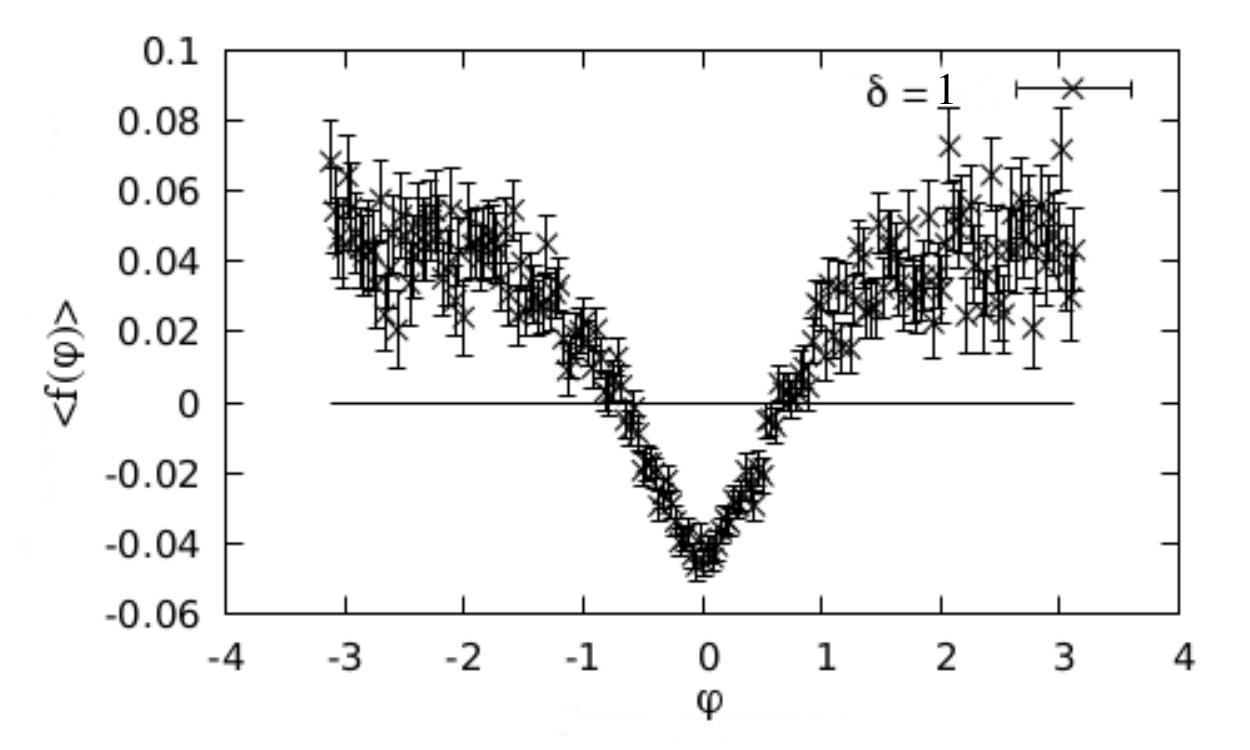}} 
\end{minipage}
\hfill
\begin{minipage}[h]{0.47\linewidth}
\center{\includegraphics[width=1\linewidth]{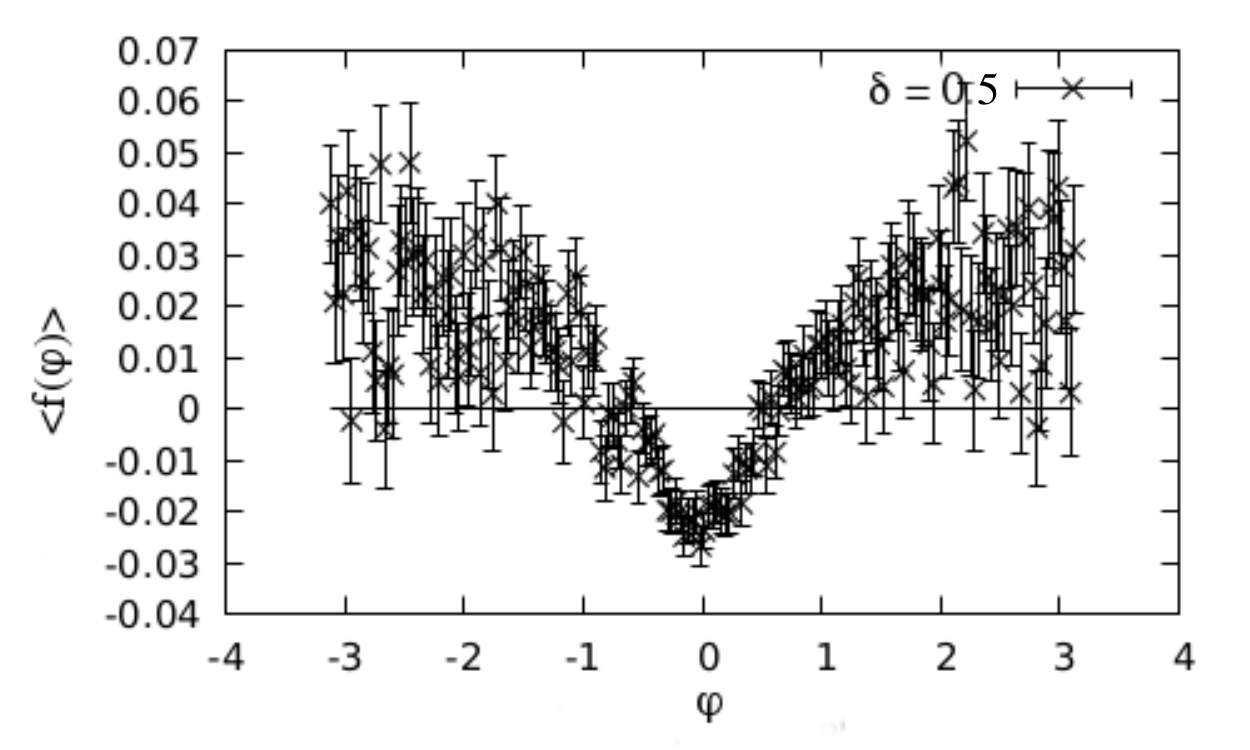}} 
\end{minipage}
\hfill
\begin{minipage}[h]{1\linewidth}
\center{\includegraphics[width=0.47\linewidth]{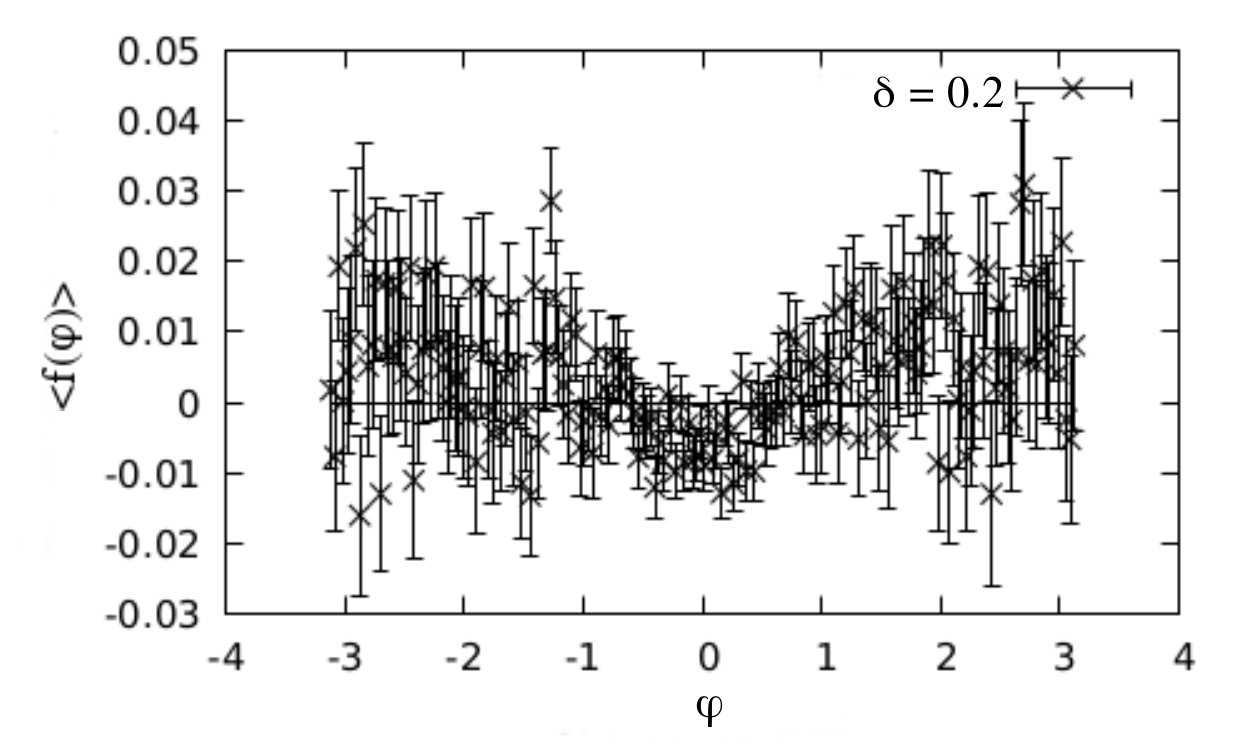}} 
\end{minipage}
\hfill
\caption{Deviation $\left<f(\phi)\right>$ for the length of the random walk step $\delta=1$, $\delta=0.5$ and $\delta=0.2$ (marked in the figures) averaged over $M=100$ groups of runs, see Exprs.~(\ref{eq:f-average},\ref{eq:f-error}).}
\label{fig:ex3}
\end{figure}

Calculation of $\left<f(\phi)\right>$ was done for $M=100$ independent runs for $\delta=1$, $\delta=0.5$ and $\delta=0.2$, and is shown in Fig.~\ref{fig:ex3}. 
Deviation is clearly depend on the size of the random walk step $\delta$, and decreases with the decreasing $\delta$. 

\begin{figure}[!ht]
\begin{center}
\includegraphics[scale=0.4]{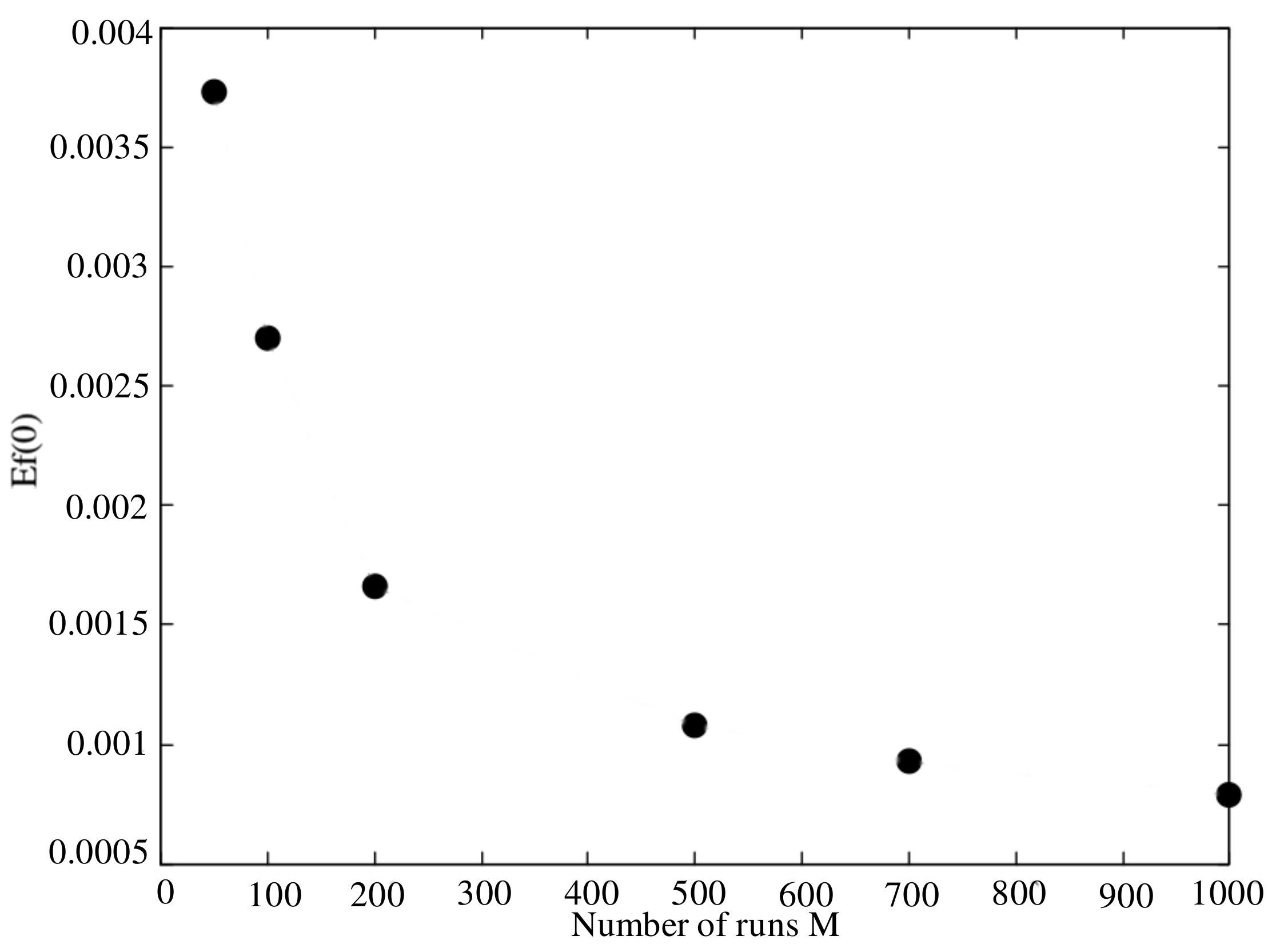}
\caption{$Ef(0)$ dependence on the number of groups $M$ of the runs.}
\label{fig:ex4}
\end{center}
\end{figure}

$P_{exp}(\phi)$ has both random and systematic error. Function $<f(\phi)>$ shows systematic component while $Ef(\phi)$ shows random component.
$Ef(\phi)$ depends on the total number of runs and decreases as $1/\sqrt{M}$ (see Fig.~\ref{fig:ex4}). 
Systematic deviation $<f(\phi)>$ depends mainly on the random walk step length $\delta$ and does not tend to 0 as $M$ grows.

\begin{figure}[!ht]
\begin{center}
\includegraphics[scale=0.4]{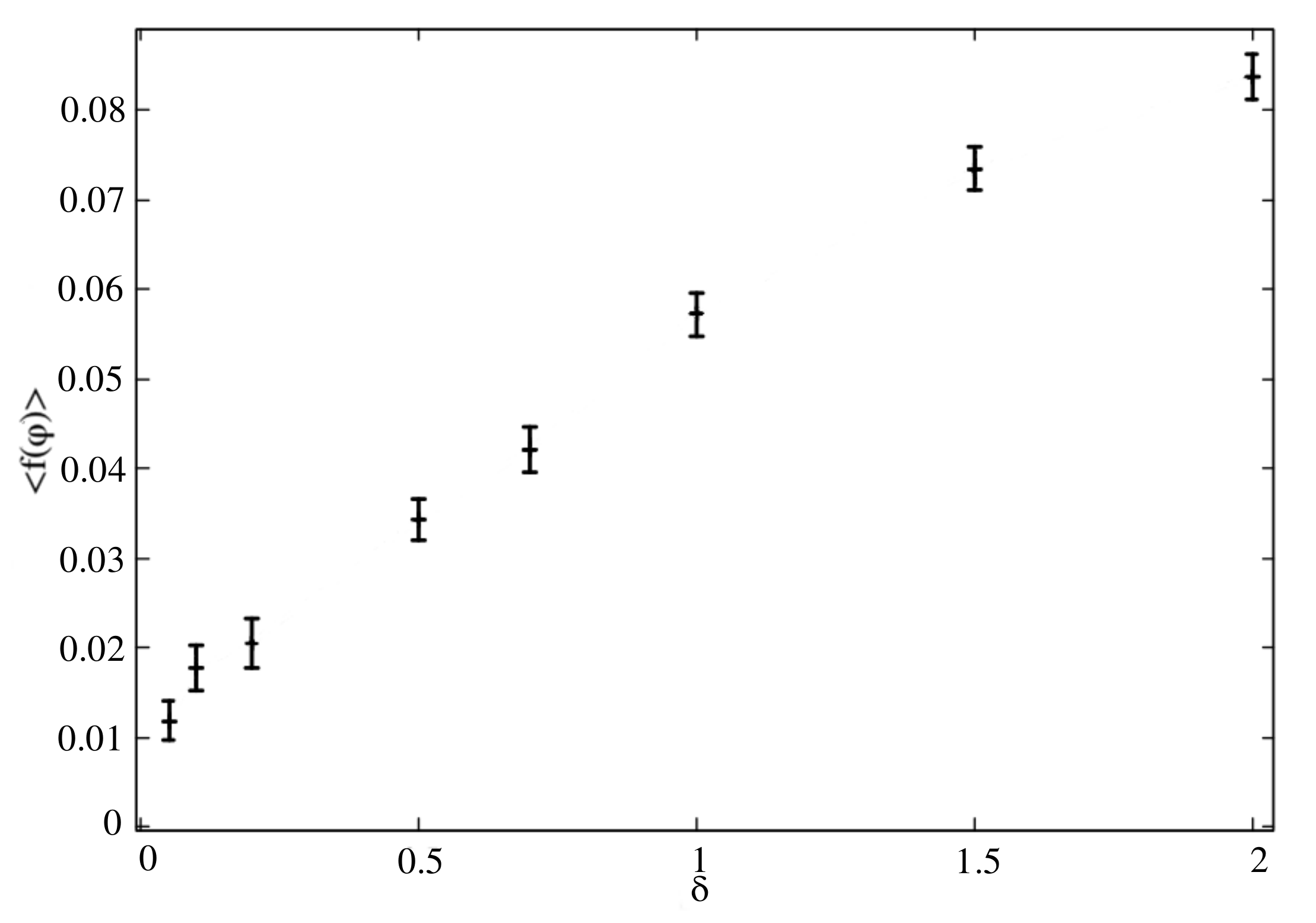}
\caption{Dependence of $<f(0)>$ on the $\delta$ for $M=100$ group of runs.}
\label{fig:ex5}
\end{center}
\end{figure}

For the reason of clarity we restrict our analysis to a single value $<f(0)>$  of $<f(\phi)>$ at $\phi=0$ 
as a measure of deviation of $P_{exp}$ from $P_{exact}$.
Fig.~\ref{fig:ex5} shows $<f(0)>$ as a function of random walk step length $\delta$. This function shows decrease when $\delta$
decreases, but the way it converges to zero could not be derived from the available data.
As $\delta$ goes to 0 various errors could become significant, e.g. rounding errors during calculation,  and special care should
be taken. We leave this question open for the future research. 

\section{Discussions}

In this paper we have estimated numerically systematic deviation of the first-passage probability for the random walking particle to hit circle of radius $R$ in the plane.  
We may conclude from the results of our simulations, and especially from the analysis of the data presented in the Figures~\ref{fig:ex3},\ref{fig:ex5},  
that the following form of the correction may take place
\begin{equation}
P_{exp}(\phi)\approx P(\phi)\left(1-\left(\frac{\delta}{R}\right)^\alpha cos(\phi)\right).
\label{eq:proposal}
\end{equation}
We propose this expression is the first-order correction to the exact result of the Laplace solution~(\ref{eq-return}).

Future work, both simulations and analytical study, have to be done to check our proposal. Our findings can be important not only for the simulation
of random walk in plane, but also in the study of boundary value problems, as well as in random growth fractals sumulations, e.g. in DLA problem~\cite{M-scaling}.

\ack
This work was supported by grant 14-21-00158 from the Russian Science Foundation.

\end{document}